\documentclass{article}
\usepackage{ijcai19}

\usepackage{times}
\usepackage{soul}
\usepackage{url}
\usepackage[hidelinks]{hyperref}
\usepackage{color}
\usepackage[utf8]{inputenc}
\usepackage[small]{caption}
\usepackage{graphicx}
\usepackage{amsmath}
\usepackage{booktabs}
\usepackage{cite}
\usepackage{comment}
\usepackage{amsmath,amssymb,amsfonts}
\usepackage[algo2e,linesnumbered,lined,boxed,commentsnumbered,ruled]{algorithm2e}
\usepackage{graphicx}
\usepackage{textcomp}
\usepackage{xcolor}
\usepackage{tikz}
\usepackage{makecell}

\newtheorem{proposition}{Proposition}

\newtheorem{definition}{Definition}

\newtheorem{remark}{Remark}

\title{Towards Automatic  Linearization via SMT Solving}

\author{
Jian Cao$^1$\and
Liyong Lin$^1$\footnote{Contact Author}  \and Lele Li$^1$ \\
\affiliations
$^1$Contemporary Amperex Technology Limited\\
\emails
llin5@e.ntu.edu.sg
}

\begin{document}

\maketitle

\begin{abstract}
Mathematical optimization is ubiquitous in modern applications. 
However, in practice, we often need to use nonlinear optimization models, for which the existing optimization tools such as Cplex or Gurobi may not be directly applicable and an (error-prone) manual transformation  often has to be done. Thus, to address this issue, in this paper we investigate the problem of automatically verifying and synthesizing reductions, the solution of which may allow an automatic linearization of nonlinear models. We show that the synthesis of reductions can be formulated as an $\exists^* \forall^*$ synthesis problem, which can be solved by an SMT solver via the counter-example guided inductive synthesis approach (CEGIS). 
\end{abstract}

\section{Introduction}
Mathematical optimization is among the most important technological backbones of many modern applications in finance, engineering, supply chain management and so on. Being able to correctly formulate an optimization model for a given application is crucial in practice, an art that might require substantial training to master for most   beginners, and it can be an error-prone process even for well-trained engineers. In practice, however, the most intuitive optimization models for the applications, such as scheduling and production planning, often involve logical or nonlinear operations, and these are often the models that are derived in the first step by optimization engineers. After this nonlinear optimization model is derived, then the engineers may proceed to transform or relax it into a linear formulation, if possible. 

In the above-mentioned process, to ensure a correct mathematical optimization model is (correctly) solved, we need to ensure three aspects:
\begin{enumerate}
    \item [a)] The “raw“ optimization model, i.e., the nonlinear optimization model before the transformation is carried out, correctly captures the user's intent and models the problem to be solved.
    \item [b)] The transformed model is equivalent to the raw model.
    \item [c)] The solver is correct in solving the transformed model.
\end{enumerate}
It is relatively difficult to ensure a), since one needs to ensure the raw optimization model is semantically an abstraction of the problem to be solved and the solution of the model can be faithfully executed, discarding irrelevant details.  It is tedious, but still technically possible, to ensure c)\footnote{There are generally two approaches, either by proving the solver itself to be correct in solving every instance or  providing a certificate for any given instance solved by the solver.}. Both a) and c) are not the focus of this work, however.  Usually, b) is performed by experienced optimization engineer by manual transformation, using tactics that are mastered by either learning or trial-and-error experiments. However, such a  manual transformation is not always reliable, since errors could be introduced in the transformation process that there is no guarantee of model equivalence or correctness, and a rigorous proof is often lacking. This work's main focus is to address b).

More abstractly, we are interested in the following technical problems in the long run. 
\begin{enumerate}
    \item [1)] {\bf Verification}: Given an (optimization) model $\Sigma$ from a class $\mathcal{C}$ and another model $\Sigma'$ from a class $S$, how shall one automatically (dis)-prove their equivalence? 
    \item [2)] {\bf Exact Synthesis}: Given a model $\Sigma$ from a class $\mathcal{C}$, how to synthesize an equivalent model $\Sigma'$ in a different class $\mathcal{S}$ or prove that there is no such equivalent model?
    \item [2)] {\bf Approximate Synthesis}: Given a model $\Sigma$ from a class $\mathcal{C}$, how shall one synthesize a “close” model $\Sigma'$ in a different class $\mathcal{S}$?
\end{enumerate}
For example, class $\mathcal{C}$ can represent the class of nonlinear programming models, while class $S$ could represent the class of linear programming models. We remark that, of course, more delicate distinctions between classes can also be made by allowing different sets of  nonlinear functions to extend the base optimization classes.  
For example, the class of mixed integer linear programming (MILP) models, when it is extended with the {\it max} function, can be  shown to be expressively equivalent to itself.

The two classes $\mathcal{C}$ and $S$ need not be different, however. In principle, the above formulation of the problem permits one to derive an efficient model from a less efficient one, despite both being in effectively the same class of models. As a well known example, one can make an MILP model more efficient by using a good formulation~\cite{Hooker} or by adding cuts\cite{TAF20}. For the problem of approximate synthesis, one is often interested in finding approximation $\Sigma'$ in a more efficient class of models that offers a good bound for a given model $\Sigma$.

$\Sigma'$ can be viewed as a (candidate) reduction of $\Sigma$. Thus, the above three problems could be viewed as the verification and synthesis of reductions between (the decision versions of) the two classes of  optimization problems (equipped with appropriate signatures), similar in spirit to~\cite{L17}. Without restriction, it is clear that all the above three problems are undecidable in general. 

In this paper, we will mainly look at the problems of verification and exact synthesis. In practice, there are often bounds on the variables that appear in the  optimization model. Thus, instead of solving the unbounded verification/synthesis problem, we focus on a bounded approach~\cite{HZWW24, FS13}. 
In addition, for  bounded synthesis, we shall also  parameterize the complexity of candidate reductions by restricting\footnote{The details will be explained later.} the numbers of auxiliary variables and  linear inequalities,  enabling 
an incremental search~\cite{b1, b2} of bounded reductions. As we shall see later, 
the method is quite effective for a class of nonlinear models that are essentially piece-wise linear\footnote{For example, the nonlinear function {\it max} is piece-wise linear.}, where automated reasoning tools such as SMT solvers~\cite{MB08, cvc} can be used for automated reduction verification/synthesis. 



The main contributions of this work are listed as follows:
\begin{itemize}
    \item An approach is given to reduce the  problem of bounded  verification into solving SMT formulas. We remark that, in principle, the approach can produce a counter example to witness non-equivalence of models.
    \item We show that the counter-example guided inductive synthesis (CEGIS) approach could be adopted to synthesize an optimal reduction (at least in principle), under a natural measure (i.e., the numbers of  auxiliary variables and linear inequalities).
\end{itemize}

{\bf Related Works: }
The problems of verifying and synthesizing reductions have been studied in different contexts.   \cite{B23} proposes an approach to allow users to  interactively verify the correctness of reductions from disciplined convex programs (DCP) to conic programs by using Lean, an interactive theorem prover~\cite{MU21}.
 The earliest work that we know of in synthesizing reductions, which are captured by “gadgets”, between decision problems can be found in~\cite{TSSW96}, by using linear programming techniques. 
In~\cite{b1, b2}, based on descriptive complexity theory,  quantifier free projections (a rather weak class of reductions) between complexity classes are automatically synthesized via SAT solvers. A reduction in~\cite{b1, b2}  is captured by a Boolean query, which is parameterized by the size of the structure that the reduction is supposed to be effective, as well as another two parameters that characterize the complexity of the reduction. The counter-example guided inductive synthesis (CEGIS) approach is used to address the $\exists^* \forall^*$ synthesis problem. Later, the work has also been extended to accommodate machine learning techniques~\cite{C16}.  In a different context, 
\cite{L17, L17ACC, L18}
work on the problem of automatically verifying and synthesizing reductions for a particular class of verification problems, which allows automatic parallel verification and a substantial reduction of verification complexity. The technique of automaticaly computing a reduction can be used to obtain a compact MILP model encoding of neural networks with layers that use piecewise-linear functions such as ReLU or
max-pooling~\cite{TXT17}. This MILP encoding  can then be used for downstream neural network verification tasks~\cite{TXT17}.

\section{Verification and Synthesis of Reductions}
In this section, we first provide a motivating example to illustrate the idea behind the verification and synthesis approach. 
\subsection{Motivating Examples}

{\bf Nonlinear Constraints:}

For the nonlinear constraint $c=max\{a, b\}$, the following
linearization 

\begin{equation}
\begin{cases} 
    c \geq a\\
    c \geq b\\
    c \leq a+(1-u_1)M \\
    c \leq b+(1-u_2)M\\
    u_1+u_2 \geq 1
\end{cases}
\end{equation}
is well known, where $u_1, u_2 \in \{0, 1\}$ are two newly introduced Boolean variables, and $M$ is a large constant. 

Suppose the nonlinear constraint $c=max\{a, b\}$ appears in an optimization model, then it can be replaced by the above linearization reduction. In practice, however, to safely use the above  linearization reduction, we have to show its correctness, which amounts to establishing the following. 
\begin{proposition} Let $M > |a|+|b|+|c|+1$. Then, for any $a, b, c \in R$, $c=max\{a, b\}$ iff  \\
\begin{equation}
\exists u_1, u_2 \in \{0,1\},
\begin{cases} 
    c \geq a\\
    c \geq b\\ 
    c \leq a+(1-u_1)M \\
    c \leq b+(1-u_2)M\\
    u_1+u_2 \geq 1
\end{cases}
\end{equation}
\end{proposition}
{\em Proof}: Suppose $c=max\{a, b\}$, then it is clear that $c \geq a$ and $c \geq b$. Also, $c \leq a$ or $c \leq b$ holds by the definition of $max$. Suppose $c \leq a$, then we can set $u_1=1$ and $u_2=0$, and it holds that $c \leq a+(1-u_1)M$, $c \leq b+(1-u_2)M$ and $u_1+u_2 \geq 1$. On the other hand, suppose $c \leq b$, then we can set $u_1=0$ and $u_2=1$ and it holds that $c \leq a+(1-u_1)M$, $c \leq b+(1-u_2)M$ and $u_1+u_2 \geq 1$.

Suppose there exists some $u_1, u_2 \in \{0, 1\}$ such that $c \geq a$, $c \geq b$, $c \leq a+(1-u_1)M$, $c \leq b+(1-u_2)M$ and $u_1+u_2 \geq 1$. Any such a choice of $u_1, u_2$ must satisfy the requirement that $u_1=1$ or $u_2=1$. Without loss of generality, suppose $u_1=1$. Then, by $c \leq a+(1-u_1)M$ we have $c \leq a$. Then, we have $a=c \geq b$ and thus $c=max\{a, b\}$.

{\bf Nonlinear Objectives:}

Consider the following nonlinear program, where the feasible region $D$ is assumed to be linear.

\begin{align}
    &\text{min} && \text{max}\{a, b\} \\ 
    &\text{s.t.} && (a,b) \in D
\end{align}

It is straightforward to see that the above program is equivalent to the following, where $z$ is a newly introduced auxiliary variable. 

\begin{align}
    &\text{min} && z \\ 
    &\text{s.t.} && (a,b) \in D \\
    &  && z=\text{max}\{a, b\}
\end{align}

Thus, without loss of generality, in the rest of this work we shall only consider nonlinear constraints. 

\subsection{Methodology}
In this section, we show how a predicate may be linearized\footnote{In this work, reduction is  synonymous with linearization.}. 

Given a  predicate  $\Phi(y_1, y_2, \ldots, y_m)$, where $y_i \in D_i \subseteq \mathbb{R}$ for each $i \in [1, m]$, we introduce a tuple of boolean variables $(u_1, u_2, \ldots, u_k) \in \{0,1\}^k$. Let  ${\bf y}=(y_1, y_2, \ldots, y_m)$. 
\begin{definition}
The predicate $\Phi({\bf y})$ is  $(l, k)$-linearizable iff 
\begin{center}
$\exists {\bf X} \in \mathbb{R}^{l \times (m+k+1)}, \forall {\bf y} \in D_1\times D_2 \times \ldots \times D_m$,

$\bigvee_{{\bf u} \in \{0,1\}^k}(\Phi({\bf y}) \iff {\bf X} [{\bf y}, {\bf u}, 1]^T \leq {\bf 0})$, 
\end{center}
where {\bf 0} is a tuple of 0's of size $l$.
\end{definition}
\begin{remark}
Intuitively, we look for a valuation of ${\bf X}$ such that, the region defined by the set of linear inequalities, after being projected to ${\bf y}$, spans the universe  $D_1\times D_2 \times \ldots \times D_m$. Thus, $u_1, u_2, \ldots, u_k$ act as auxiliary Boolean variables.
\end{remark}
\begin{remark}
Since the valuation of ${\bf X}$ uniquely determines the set of linear inequalities, including its dimensions, hereafter we directly refer to a valuation of ${\bf X}$ as a reduction of $\Phi({\bf y})$. 
\end{remark}
\begin{definition}
The predicate $\Phi({\bf y})$ is said to be linearizable iff 
it is $(l, k)$-linearizable for some $l \geq 1, k \geq 0$.
\end{definition}
\begin{remark}
In the above discussion, we only refer to a predicate; however, the definition of  linearization and the related verification and synthesis results, to be further discussed, can be straightforwardly extended to a set of predicates. 
\end{remark}
We do not distinguish between a variable and its valuation in the rest of this work. In fact, only variables will be quantified. From Remark 2, a counter example to a candidate reduction ${\bf X} \in \mathbb{R}^{l \times (m+k+1)}$ is a valuation ${\bf y}  \in D_1\times D_2 \times \ldots \times D_m$ such that
\begin{center}
$\bigwedge_{{\bf u} \in \{0,1\}^k}((\Phi({\bf y}) \wedge (\bigvee_{i=1}^l {\bf X}[i:][{\bf y}, {\bf u}, 1]^T >0)) \vee (\neg \Phi({\bf y}) \wedge {\bf X}[{\bf y}, {\bf u}, 1]^T \leq 0))$
\end{center}
holds. It is clear that the problem of finding counter examples for a candidate reduction ${\bf X} \in \mathbb{R}^{l \times (m+k+1)}$ can be delegated to a SMT solver. In the rest of this section, we shall explain how the counter-example guided inductive synthesis (CEGIS) approach can be used to synthesize reductions. 

The idea of the CEGIS approach for synthesizing an $(l, k)$-linearization is explained as follows. A set $S=\{{{\bf y_1}, \ldots, {\bf y_n}}\}$ of valuations is initialized and a candidate reduction ${\bf X}$ is synthesized to work for $S$ by solving Formula 1.

{\bf Formula 1:  Reduction Finding}
\begin{center}
$\exists {\bf X} \in \mathbb{R}^{l \times (m+k+1)}, \bigwedge_{{\bf y} \in S} \bigvee_{{\bf u} \in \{0,1\}^k}(\Phi({\bf y}) \iff {\bf X} [{\bf y}, {\bf u}, 1]^T \leq {\bf 0})$  
\end{center}
If no such an ${\bf X}$ exists, then we can set $l:=l+1, k:=k+1$ and repeat solving Formula 1 for the updated $l$ and $k$; else, let ${\bf X}$ be obtained, then we know that ${\bf X}$ is a candidate reduction that works for $S \subseteq D_1\times D_2 \times \ldots \times D_m$; however, we are still not yet certain ${\bf X}$ is a reduction, as it is not fully verified on $D_1\times D_2 \times \ldots \times D_m$. Indeed, we need to  solve the problem of finding a counter example for ${\bf X}$, i.e., solving Formula 2.

{\bf Formula 2: Reduction Refutation} 
\begin{center}
$\exists {\bf y}\in D_1\times D_2 \times \ldots \times D_m$, \\ $\bigwedge_{{\bf u} \in \{0,1\}^k}((\Phi({\bf y}) \wedge (\bigvee_{i=1}^l {\bf X}[i:][{\bf y}, {\bf u}, 1]^T >0)) \vee (\neg \Phi({\bf y}) \wedge {\bf X}[{\bf y}, {\bf u}, 1]^T \leq 0))$
\end{center}
If no such a ${\bf y}$ exists, then ${\bf X}$ is verified to be a reduction, since no counter-example could be found; else, let ${\bf y}$ be obtained, then we know that ${\bf y}$ is a counter example that works for ${\bf X}$ (that is, ${\bf X}$ is refuted by ${\bf y}$) and it is clear that ${\bf y} \notin S$. We then can set $S:=S \cup \{y\}$ and repeat solving Formula 1 for the updated $S$ to find a new candidate reduction ${\bf X}$.

\begin{remark}
With a minor modification, the above incremental approach is able to find an optimal reduction in principle, if such a reduction indeed exists.  
\end{remark}
\section{Conclusions}
We have presented an approach to verify and synthesize reductions by using SMT solvers. As a work-in-progress, there is an obvious weakness of our approach, that is, the scalability to large instances (including unbounded synthesis) is limited. As the expressive power of MILP models can be restrictive, a viable path is to use our approach for approximate synthesis, where much larger class of nonlinear models can be dealt with. These issues will be studied in future works.

\section*{Acknowledgement}
This work is financially supported by National Key R\&D Program of China (Grant 2022YFB4702400).


\begin{thebibliography}{00}
\bibitem{Hooker}
J. Hooker, ``Formulating good MILP models'', Wiley Encyclopedia of Operations Research and Management Science, 2011.

\bibitem{TAF20}
Y. Tang, S. Agrawal, Y. Faenza, “Reinforcement learning for integer programming: learning to cut'', Proceedings of the 37th International Conference on Machine Learning, 119:9367-9376, 2020.

\bibitem{L17}
L. Lin, T. Masopust, W. M. Wonham and R. Su, "Automatic generation of optimal reductions of distributions," IEEE Transactions on Automatic Control, vol. 64, no. 3, pp. 896-911, 2019.

\bibitem{HZWW24}
Y. He, P. Zhao, X. Wang, Y. Wang, “VeriEQL: bounded equivalence verification for complex SQL queries with integrity constraints”, Proceedings of the ACM on Programming Languages 8: 1071 - 1099, 2024.

\bibitem{FS13}
B. Finkbeiner, S. Schewe, “Bounded synthesis”, International Journal on Software Tools for Technology Transfer,  15, 519–539, 2013.

\bibitem{b1} M. Crouch, N. Immerman, J. Eliot B. Moss, ``Finding reduction automatically'',  Fields of Logic and Computation, pp. 181-200, 2010.

\bibitem{b2} C. Jordan, Ł. Kaiser. ``Experiments with reduction finding'', International
Conference on Theory and Applications of Satisfiability Testing, pp.
192-207, 2013.

\bibitem{MB08}
 L. de Moura, N.  Bjørner, “Z3: An efficient SMT solver”, International Conference on Tools and Algorithms for the Construction and Analysis of Systems, 2008.

 \bibitem{cvc}
 H. Barbosa et al., “cvc5: A versatile and industrial-strength SMT solver”, International Conference on Tools and Algorithms for the Construction and Analysis of Systems, 2022.
 
\bibitem{B23}
A. Bentkamp, R. Fernández Mir, J. Avigad,
``Verified reductions for optimization'', International Conference on Tools and Algorithms for the Construction and Analysis of Systems
, 74-92, 2023.

\bibitem{MU21}
L.d. Moura, S. 
Ullrich, ``The Lean 4 theorem prover and programming language'', In: Automated Deduction–CADE 28, 2021.

\bibitem{TSSW96}
L. Trevisan, G.B. Sorkin, M. Sudan, D.P. Williamson,
``Gadgets, approximation, and linear programming'', 
Proceedings of 37th Conference on Foundations of Computer Science,  pp. 617-626, 1996. 


\bibitem{C16}
C. Jordan, L. Kaiser,
``Machine learning with guarantees using descriptive complexity and SMT solvers'', CoRR abs/1609.02664, 2016.


\bibitem{L17ACC}
L. Lin, S. Ware, R. Su, W. M. Wonham, ``Reduction of distributions and
its applications'', American Control Conference, pp. 3848-3853, 2017.

\bibitem{L18}
L. Lin, T. Masopust, W. M. Wonham, R. Su,
``Automatic generation of optimal reductions of distributions'', IEEE Trans. of Automatic
Control, 62(11): 5755-5768, 2017.






\bibitem{TXT17}
V. Tjeng,  K. Y. Xiao, R. Tedrake,  “Evaluating robustness of neural networks with mixed integer programming.” International Conference on Learning Representations, 2017.




\end{thebibliography}
\end{document}